\documentclass[]{aa} 

\usepackage{graphicx,hyperref}
\usepackage{txfonts}
\usepackage{xcolor}

\newcommand{\tar}{J1430+2303}

\begin{document} 

\title{A Delayed Radio Flare Traces Kinetic Energy Injection in the SMBHB Candidate SDSS~J143016.05+230344.4}
\titlerunning{A delayed radio flare in J1430+2303}
   \author{Tao An
          \inst{1}
         \and 
         Ailing Wang\inst{2,3}
          \and
         Yingkang Zhang\inst{1}
         \and 
         Lei Yang\inst{4}
         \and 
         Xinwen Shu\inst{4}
         \and 
         Fabao Zhang\inst{4}
         \and
         Ning Jiang\inst{5}
         \and
         Tinggui Wang\inst{5}
         \and
         Huan Yang\inst{6}
         \and
         Zhen Pan\inst{7}
         \and
         Liming Dou\inst{8}
        \and 
        Zhijun Xu\inst{1}
         \and
         Zhenya Zheng\inst{1}
         \and 
         Ruiqiu Lin\inst{9}
         \and
         Xiaofeng Li\inst{10}
          }

   \institute{Shanghai Astronomical Observatory, Chinese Academy of Sciences, Nandan Road 80, Shanghai 200030, China  \email{antao@shao.ac.cn}
    \and Spallation Neutron Source Science Center, 523803 Dongguan, China 
    \and 
    Key Laboratory of Particle Astrophysics, Institute of High Energy Physics, Chinese Academy of Sciences, Beijing 100049, China 
      \and   Department of Physics, Anhui Normal University, Wuhu, Anhui, 241002, China \email{xwshu@ahnu.edu.cn}      
    \and CAS Key Laboratory for Researches in Galaxies and Cosmology, Department of Astronomy, 
      University of Science and Technology of China, Hefei, Anhui 230026, China 
      \and Department of Astronomy, Tsinghua University, Beijing 100084, China 
      \and Tsung-Dao Lee Institute, Shanghai Jiao-Tong University, Shanghai, 520 Shengrong Road, 201210, China 
        \and
            Department of Astronomy, Guangzhou University, Guangzhou 510006, China 
        \and University of Massachusetts Amherst, 710 North Pleasant Street, Amherst, MA 01003-9305, USA 
        \and Changzhou University, Changzhou , China
}

   \date{Received ; accepted }

 
  \abstract
   {SDSS~J143016.05+230344.4 ($z=0.08105$) has been proposed as a candidate pre-coalescence supermassive black hole binary and shows remarkable multiwavelength variability. Its radio evolution provides a direct probe of the compact emitting region and of the physical origin of the late-time activity. }
   {We aim to localize the variable radio emission, characterize its spectral evolution, and constrain whether the radio brightening is produced by a newly emerging compact component, external absorption, or dissipation in a structured circumnuclear environment.}
   {We analyzed 4.7--22.2 GHz Very Long Baseline Interferometry (VLBI) observations obtained between 2022 February and 2024 February, together with quasi-simultaneous connected-array spectra covering 0.7--16.5 GHz. We combined the VLBI brightness-temperature and compactness constraints with spectral decomposition and equipartition-based estimates of source size and ambient density structure.}
   {At all epochs, the radio emission is dominated by a single unresolved milliarcsecond core with $T_{\rm B} \gtrsim 10^{7}$\,K, constraining the variable emission to $\lesssim 0.3$\,pc. The broadband spectra require two synchrotron self-absorbed components: a persistent low-frequency component with $\nu_{\rm p,steady} \approx 0.74$\,GHz and $S_{\rm p,steady} \approx 1.22$\,mJy, and a flare component whose turnover evolves from $(6.35\,{\rm GHz}, 0.18\,{\rm mJy})$ in 2022 February--May to $(8.61\,{\rm GHz}, 0.38\,{\rm mJy})$ in 2022 December, and then to $(5.83\,{\rm GHz}, 0.25\,{\rm mJy})$ in 2023 March--April. The flare contribution at 15\,GHz reaches $\sim 80\%$ and matches the near-epoch VLBI recovery fraction, showing that the high-frequency brightening arises from a newly formed compact synchrotron component. A second brightening of the 15.2\,GHz VLBI core is detected between 2023 September and 2024 February, while the source remains unresolved. Equipartition scalings imply characteristic radii of $\sim 5 \times 10^{-4}$\,pc for the flare and $\sim 9 \times 10^{-3}$\,pc for the steady component, and indicate a steep inner circumnuclear density profile, $n \propto R^{-1.7}$.}
   {The delayed radio flare is best explained by dissipation in an outflow or jet-base disturbance propagating through a structured circumnuclear medium. }


   \maketitle
%

\section{Introduction} \label{sec:intro}
 
Supermassive black hole binaries (SMBHBs) on the verge of coalescence are laboratories for studying non-stationary accretion: gas streams, shocks, and magnetic flux are forced to respond on dynamical timescales, while gravitational radiation steadily drains angular momentum.  This is also the electromagnetic regime most relevant for low-frequency gravitational-wave astronomy---the population targeted today by Pulsar Timing Arrays and, in the coming decade, by LISA \citep{2018ApJ...865..140D, 2018ApJ...853L..17B, 2022LRR....25....3B, 2023ApJ...951L...8A, 2023LRR....26....2A}.  In parallel, James Webb Space Telescope (JWST) is pushing direct observations of rapid black-hole growth and dense nuclear gas reservoirs to high redshift, demonstrating that mergers and violent accretion episodes should be common across cosmic time.  Characterizing the underlying physics of these systems remains inherently difficult, as the decisive physics takes place on sub-parsec scales, where most diagnostics are either unresolved or heavily absorbed.  Radio interferometry is one of the few tools that can directly localize the dissipative region and convert a radio spectrum into size, magnetic field, and energy scales.

SDSS~J143016.05+230344.4 (hereafter J1430+2303; $z=0.08105$) became prominent after a sequence of decaying-period optical/X-ray flares in early 2022, interpreted by \citet{2022arXiv220111633J} as an eccentric SMBHB approaching merger.  Subsequent work has highlighted viable single-SMBH alternatives, including stochastic red-noise processes and inner-corona instabilities; the nature of the event remains open \citep[e.g.,][]{2023MNRAS.518.4172D, 2023ApJ...945L..34M}.  Either way, the source offers something rare: a nearby nucleus that changed state on human timescales, with a well-defined high-energy trigger epoch that we can use to anchor delayed radio activity.

Radio emission provides a kinetic energy tracer complementary to optical/X-ray diagnostics.  Compact synchrotron sources (a weak jet base, a plasmoid, or a shock driven by a disk wind/outflow) are expected to be self-absorbed at GHz frequencies and to brighten first at high frequency.  Very Long Baseline Interferometry (VLBI) observations can then test whether the variability is truly nuclear or instead contaminated by host-scale emission.  Early VLBI, enhanced Multi-Element Radio Linked Interferometer Network (e-MERLIN), and Australia Telescope Compact Array (ATCA) observations established that J1430+2303 is radio-quiet on kiloparsec scales but hosts a faint, flat-spectrum compact core \citep[][and references therein]{2022ATel15306....1B, 2022ATel15267....1A, 2022A&A...663A.139A}.  What was missing is a time-resolved view of the post-outburst radio evolution: whether a new compact component emerges, how its synchrotron turnover migrates, and what that implies for particle acceleration and the circumnuclear medium (CNM).  This is exactly the kind of empirical template that Rubin/LSST, Euclid, and Roman will increasingly demand as they deliver more rapidly varying nuclei and candidate sub-parsec SMBHBs.

In this paper we combine multi-epoch 4.7--22.2\,GHz VLBI imaging (2022--2024) with quasi-simultaneous 0.7--16.5\,GHz connected-array spectra.  We introduce a time-resolved \emph{steady+flare} decomposition of the broadband spectrum across three epochs and map the resulting spectrum turnover evolution onto physical constraints on the size, magnetic field, and ambient medium density.  Throughout we adopt a flat $\Lambda$CDM cosmology with $H_0=70$~km~s$^{-1}$~Mpc$^{-1}$ and $\Omega_M=0.3$, giving $D_L\simeq368$~Mpc and 1~mas $\simeq1.5$~pc at $z=0.08105$.


\section{Observations and data reduction} \label{sec:obs}

\subsection{VLBI observations and data reduction}

We conducted new VLBI observations with Very Long Baseline Array (VLBA, BA157 from 2022 April 26 to May 31 and BA166 from 2023 September 1 to 2024 February 18) and European VLBI Network (EVN, RA006 on 2023 February 16). We also included earlier 2022 February--March VLBI epochs (EVN RS005; VLBA BA154;  \citealt{2022A&A...663A.139A}) (Table \ref{tab:VLBIobs}). Details of station participation, correlator setup and observing sequence are provided in Appendix \ref{app:obs}.

Calibration followed our automated workflow \citep{2022SCPMA..6529501A} with standard amplitude (system temperature and gain curves), bandpass, and phase referencing (Appendix \ref{app:obs}). 
Session BA157b used a broad C band split into two sub-bands centred at 4.7 and 7.6~GHz to enable an in-band spectral-index measurement; the sub-bands were calibrated and imaged separately and, for display, their visibilities were also concatenated to form a 6.2~GHz continuum dataset. Given the faintness of the target at all observed frequencies, images were produced using conservative deconvolution and strictly without self-calibration.

We parameterized the compact emission by fitting a single circular Gaussian directly to the visibilities using the \textsc{modelfit} program in \textsc{Difmap}. For fits that converge to a point source due to extreme compactness, we conservatively adopt the theoretical minimum resolvable size \citep{2005astro.ph..3225L} as an upper limit on the FWHM. Total flux-density uncertainties are computed as the quadrature sum of the statistical image \textit{rms} and the systematic visibility-amplitude scale error, derived from system temperatures, gain-curve accuracy, and bandpass stability.
To robustly validate these parameters, we performed an independent visibility-domain Markov chain Monte Carlo (MCMC) analysis to sample the joint posteriors of flux density, size, and position (Appendix~\ref{app:mcmc}). While the \textsc{modelfit} and \textsc{MCMC} flux densities are statistically consistent across all bands and epochs, the Bayesian approach intrinsically handles parameter degeneracies better, yielding tighter and more reliable morphological constraints. Consequently, we adopt the MCMC-derived posterior medians (68\% credible intervals) and 95\% upper limits for unresolved cases as our fiducial measurements in Table~\ref{tab:mod}.

\subsection{VLA and uGMRT observations} \label{sec:connected}

We obtained three Karl G. Jansky Very Large Array (VLA) epochs bracketing the early-2022 high-energy activity. \emph{Epoch~I} (A configuration; 2022 Apr~6--May~6; project code: 21B-375) covered S/C/X/Ku bands. 
Complex gains were tied to the phase calibrators J1436+2321 (S/C/X) and J1407+2827 (Ku); 3C\,286 set the flux density scale and was used to calibrate bandpass. The total integration time is $\sim$26 min per band. 
Data were processed using the Common Astronomy Software Applications (\textsc{CASA}, version 5.3.0) with the standard  pipeline (version 5.3.1). The calibrated data were imaged with \textsc{CLEAN} program with Briggs parameter ${\tt robust}=0$. The final images have a typical synthesized beam of 0\farcs6$\times$0\farcs5, 0\farcs3$\times$0\farcs5, 0\farcs3$\times$0\farcs2, 0\farcs2$\times$0\farcs1 at S, C, X, and Ku-band, respectively. The target is clearly detected and unresolved in all bands. The integrated-to-peak ratios lie within $\approx$0.98--1.08 (median 1.04). \emph{Epoch~II} (C configuration; 2022 Dec~26--27; project code: 22B-106) repeated S/C/Ku with the same calibrators and pipeline.  \emph{Epoch~III} (B configuration; 2023 March~7--April 26; project code: 22B-106) repeated C/X/Ku with the same calibrators and pipeline. The source remained unresolved with integrated-to-peak ratios 0.96--1.12 (median 1.01). For consistency across configurations, we adopt peak flux densities for the spectrum analysis; the observing log and measurements are listed in Table~\ref{tab:vla_gmrt}.

Quasi-simultaneous upgraded Giant Metrewave Radio Telescope (uGMRT) observations sampled the low-frequency spectrum at Band~5 (1.25\,GHz) and Band~4 (0.75\,GHz), each over 0.4\,GHz bandwidth. 3C\,286 was used to calibrate the flux density scale, and the quasar J1407+284 for calibrating complex gains. We used \texttt{tclean} for imaging with MS--MFS (multi-scale multi-frequency synthesis; two Taylor terms, \texttt{nterms}=2). Phase-only self-calibration improved the dynamic range of the image. The source remains unresolved at both bands (typical synthesized beams $\sim$1\farcs3$\times$0\farcs9 at Band~5 and $\sim$1\farcs5$\times$0\farcs7 at Band~4) with stable flux densities across the two epochs (Table~\ref{tab:vla_gmrt}).

For the time-resolved spectrum, we group connected-array measurements into three quasi-/near-epoch bins:
\noindent\emph{Epoch~I}: 2022 Feb--May (uGMRT March + VLA April--May),\
\emph{Epoch~II}: 2022 Dec (VLA + uGMRT),\
\emph{Epoch~III}: 2023 Mar--Apr (VLA).
These bins are designed to track months-scale evolution while keeping each spectrum sufficiently sampled.  VLBI flux densities are \emph{not} used in the synchrotron self-
absorption (SSA)  fitting, because VLBI resolves out larger-scale steep-spectrum emission; instead, VLBI is used to localize the flare and constrain compactness.

\section{Results}
 \label{sec:results}

\subsection{A single unresolved sub-parsec VLBI core}

J1430+2303 is detected in every VLBI epoch (Fig.~\ref{fig:vlbi_rep}; the full montage is shown in Appendix~\ref{app:obs}).  The images reach \textit{rms} levels of $10$--$30\,\mu$Jy~beam$^{-1}$, corresponding to peak signal-to-noise ratios of $\gtrsim5$--$9$.  At all frequencies the visibilities are well described by a single compact component; we find no statistically significant secondary peak and no resolved jet-like extension within the mapped fields.  The straightforward implication is that the (variable) radio power is generated in the nucleus on sub-parsec scales.

Visibility-domain modeling (Section~\ref{sec:obs}) places tight size limits on this core.  Updating our early 1.7\,GHz constraints \citep{2022A&A...663A.139A}, we obtain FWHM upper limits of $<0.19$\,mas ($<0.29$\,pc) at 5\,GHz and $<0.22$\,mas ($<0.37$\,pc) at 15\,GHz.  The corresponding lower limits on the brightness temperature span $T_{\rm B}\gtrsim(0.5$--$36)\times10^{7}$\,K, well in the non-thermal regime expected for radio-quiet AGN\citep[e.g.,][]{1998MNRAS.299..165B, 2004ApJ...603...42A, 2022ApJ...936...73A, 2023MNRAS.518...39W, 2023MNRAS.525.6064W, 2025ApJS..277...56C} and difficult to reconcile with purely thermal star-formation processes.  The relative astrometry is consistent with a stationary centroid (Fig.~\ref{fig:positions}).  Over the $\Delta t_{\rm obs}\simeq1.7$\,yr baseline, the absence of a separated compact knot implies an illustrative apparent-speed limit of $\lesssim0.5$--$0.8\, c$ (Appendix~\ref{app:speed}); the long-baseline size evolution yields an even tighter constraint, $\lesssim0.16\,c$, if interpreted as unresolved expansion.

\begin{figure}
  \centering
  \includegraphics[width=0.9\columnwidth]{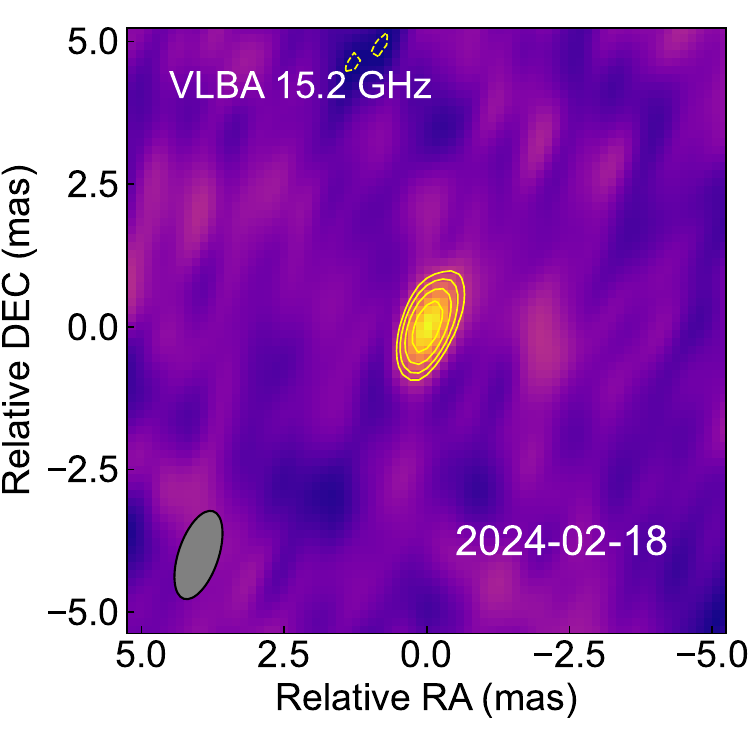} 
  \caption{Representative natural-weighted VLBI image of J1430$+$2303 at 15.2~GHz (2024 February 18; VLBA).  The restoring beam ($1.2 \, \text{mas} \times 0.5 \, \text{mas}$) is shown in the lower-left corner. The \textit{rms} noise is 0.027 mJy~beam$^{-1}$. Contours start at $3\sigma$ and increase by factors of $\sqrt{2}$. Figure~\ref{fig:tar} in Appendix presents the complete montage.}
  \label{fig:vlbi_rep}
\end{figure}

\subsection{High-frequency brightening originates in the VLBI core}

The radio evolution separates cleanly into a stable low-frequency component and a variable high-frequency component (Fig. \ref{fig:sed}).  At 5\,GHz the VLBA flux densities are consistent with being constant over the 55-day baseline in early 2022, indicating a persistent compact core.  At higher frequency the behavior changes: the VLA shows a factor of $\sim1.6$ increase at 15\,GHz from 2022 April--May to 2022 December, and the EVN detects a bright 22.2\,GHz core ($0.34\pm0.08$\,mJy) in 2023 February, confirming that a high-frequency flare lasting into early 2023.

A key cross-check comes from a near-simultaneous comparison of the VLA (2022 May 6) and VLBA (2022 May 8) at 15\,GHz: $\sim76\%$ of the connected-array flux density is recovered on milliarcsecond (mas) baselines.  This agreement localizes the bulk of the high-frequency emission including the flare component to the unresolved VLBI core.  The later 15.2\,GHz VLBA campaign further reveals renewed brightening, from $0.154$\,mJy (2023 September 1) to $0.305$\,mJy (2024 February 18), again without resolving any extended structure.  Taken together, the data point to episodic energy injection and particle acceleration operating well inside the central parsec.

\subsection{Steady+flare component decomposition} \label{sec:sed}

\begin{figure*}
\centering
\includegraphics[width=\textwidth]{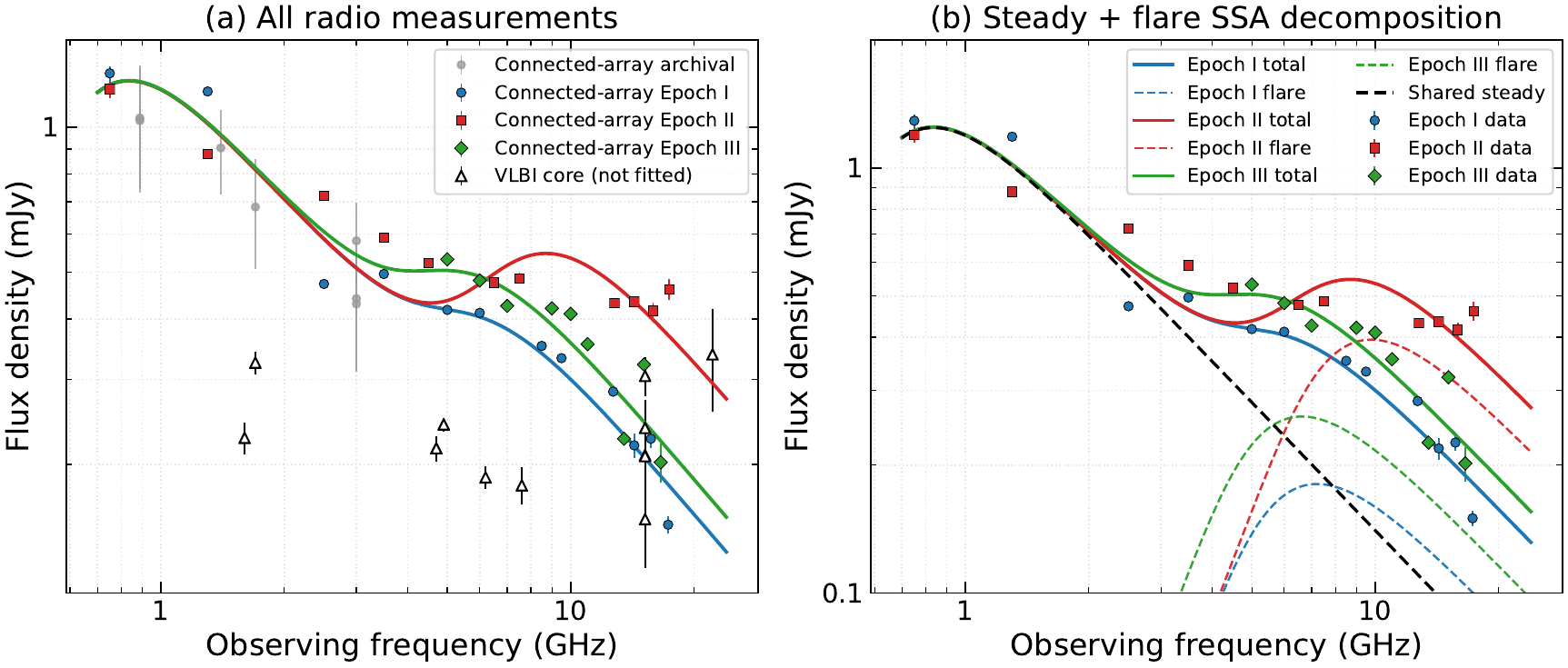}
\caption{Radio spectrum and component decomposition. \textit{Left:} all available connected-array measurements (Epochs~I/II/III; filled symbols), additional archival radio points (open symbols), and the best-fit total models for each epoch (solid curves). \textit{Right:} decomposition into a time-independent steady SSA component (black dashed curve) plus a time-variable flare SSA component (colored dotted curves). Colored solid curves show the epoch-wise totals. VLBI-core flux densities are overplotted as stars for comparison but are not used in the fit.}
\label{fig:sed}
\end{figure*}

The connected-array spectra evolve strongly between 2022 and 2023 and cannot be described by a single power law (Fig.~\ref{fig:sed}).  To isolate the variable part in a way that is minimally model-dependent, we write the total spectrum as
\begin{equation}
S_{\nu,{\rm tot}}(t)=S_{\nu,{\rm steady}}+S_{\nu,{\rm flare}}(t),
\end{equation}
and model both terms as homogeneous SSA spectra (Appendix~\ref{app:ssa}).  The steady component is constrained by the best-sampled early data and held fixed across epochs; the flare component is allowed to vary.

The best-fit steady component has $\nu_{\rm p,steady}=0.74^{+0.09}_{-0.11}$\,GHz and $S_{\rm p,steady}=1.22^{+0.12}_{-0.08}$\,mJy (Table~\ref{tab:ssa_components}).  The flare component evolves as
\begin{align}
(\nu_{\rm p,flare},S_{\rm p,flare})_{\rm I} &=
(6.346^{+0.593}_{-0.570}\,\mathrm{GHz},\ 0.176^{+0.024}_{-0.022}\,\mathrm{mJy}), \nonumber\\
(\nu_{\rm p,flare},S_{\rm p,flare})_{\rm II} &=
(8.608^{+0.851}_{-0.744}\,\mathrm{GHz},\ 0.384^{+0.027}_{-0.028}\,\mathrm{mJy}), \nonumber\\
(\nu_{\rm p,flare},S_{\rm p,flare})_{\rm III} &=
(5.831^{+0.646}_{-0.645}\,\mathrm{GHz},\ 0.253^{+0.030}_{-0.025}\,\mathrm{mJy}),
\end{align}
implying that the flare both brightened and hardened from Epoch~I to Epoch~II and then softened by Epoch~III.

At 15\,GHz, the inferred flare fraction is $\sim0.55$ (Epoch~I), $\sim0.78$ (Epoch~II), and $\sim0.62$ (Epoch~III).  The Epoch~II fraction closely matches the 15\,GHz VLBI recovery fraction, providing an independent consistency check that the variable high-frequency power is concentrated in the VLBI core.

We tested the statistical necessity of this decomposition against two alternatives: an independent one-SSA-per-epoch model and a completely non-varying single-SSA model.  The shared \textit{steady+flare} model gives an excellent fit ($\chi^2_\nu=0.52$ for 24~dof) and is strongly preferred over the independent ($\chi^2_\nu=3.51$) and non-varying ($\chi^2_\nu=4.00$) descriptions.  Information criteria also favor the \textit{steady+flare} model ($\Delta{\rm AIC}=74.8$, $\Delta{\rm BIC}=71.8$ relative to the independent-epoch case).  The flare component is detected in all three epochs at $\simeq5$--$8\, \sigma$ significance.

\section{Physical Interpretation}
\label{sec:disc}

The VLBI localization and the time-resolved spectral decomposition together reveal a simple phenomenology: a long-lived, low-frequency synchrotron component persists throughout 2022--2023, while a new, compact, self-absorbed component appears first at high frequency, brightens toward late 2022, and then fades and softens. In the framework of time-domain radio loudness, this phenomenology is naturally interpreted as the superposition of a prompt compact component and a longer-lived radio-memory component (An 2026, under review).  Below we convert that phenomenology into physical constraints on the emitting region and its environment, and we discuss what kinds of nuclear outflow can satisfy the strict compactness constraints.

\subsection{Physical scales and a steep CNM density profile} \label{sec:4.1}

For a homogeneous SSA source, the turnover pair $(\nu_{\rm p},S_{\rm p})$ provides a direct constraint on the characteristic size and magnetic field.  We adopt the usual turnover-based equipartition scalings \citep[e.g.,][]{1998ApJ...499..810C,1994ApJ...426...51R} (Appendix~\ref{app:phys}) and convert each epoch's turnover into $(R_{\rm eq},B_{\rm eq})$.  The steady component corresponds to $R_{\rm eq}\simeq9.1\times10^{-3}$\,pc and $B_{\rm eq}\simeq0.10$\,G, while the flare component remains much more compact, $R_{\rm eq}\simeq(4.2$--$5.5)\times10^{-4}$\,pc, with $B_{\rm eq}\simeq0.96$--1.35\,G (Table~\ref{tab:ssa_components}).  From Epoch~II to Epoch~III the inferred $R_{\rm eq}$ increases and $B_{\rm eq}$ decreases, consistent with an aging/expanding synchrotron region once the flare has peaked.

It is useful to turn these numbers into an energy scale.  The magnetic energy in a region of radius $R$ and filling factor $f$ is
\begin{equation}
E_B \simeq \frac{B^2}{8\pi}\,\frac{4\pi}{3}\,f\,R^3,
\end{equation}
which for the Epoch~II flare gives $E_B \simeq 4\times10^{44}\,(f/0.5)$\,erg.  If the flare traces shocked plasma and a fraction $\epsilon_B$ of the post-shock internal energy is carried by magnetic fields, the corresponding internal-energy scale is larger by $\epsilon_B^{-1}$, i.e. $E_{\rm int}\sim E_B/\epsilon_B \sim 4\times10^{45}\,(\epsilon_B/0.1)^{-1}$\,erg.  We emphasize that these are order-of-magnitude calorimetric estimates; the key point is that the flare requires a non-negligible injection of kinetic/internal energy into a region well inside $10^{-3}$\,pc.

If we interpret the flare as a shock driven into the circumnuclear medium (CNM), the magnetic field can be mapped to an ambient density.  For a strong shock (adiabatic index $5/3$), the downstream internal energy density is $u_{\rm sh}\simeq(9/8)\rho v^2$.  Writing $B^2/(8\pi)=\epsilon_B u_{\rm sh}$ yields
\begin{equation}
n \simeq \frac{B^2}{9\pi\,\epsilon_B\,m_p\,v^2}
\simeq 1.1\times10^{4}\,
\left(\frac{\epsilon_B}{0.1}\right)^{-1}
\left(\frac{v}{0.2c}\right)^{-2}\ {\rm cm^{-3}}
\end{equation}
for the Epoch~II flare field ($B_{\rm eq}\simeq1.35$\,G).  Applying the same mapping to the steady component gives $n\sim60\ {\rm cm^{-3}}$ at $R\simeq9\times10^{-3}$\,pc.  Connecting these two characteristic radii implies a steep inner profile, $n(R)\propto R^{-k}$ with
\begin{equation}
k\simeq -\frac{\ln(n_{\rm flare}/n_{\rm steady})}{\ln(R_{\rm flare}/R_{\rm steady})}\approx 1.7.
\end{equation}
Because $k$ depends mainly on ratios, it is comparatively insensitive to uncertain absolute normalizations (geometry, modest Doppler factors, or microphysical parameters that shift both components similarly). Recent studies of radio-detected TDEs have likewise shown that radio evolution can provide direct constraints on sub-parsec circumnuclear density profiles, with sample-average behavior broadly consistent with Bondi-like scaling and closure-relation analyses yielding similarly steep CNM gradients in many systems \citep{2026arXiv260214838G,2026arXiv260306056Z}. The robust inference is therefore geometric: the compact flare evolves inside a rapidly rising CNM density toward $\sim10^{-4}$--$10^{-2}$\,pc scales.

A practical caveat is whether the observed turnover could be set, in part, by external free--free absorption (FFA) rather than SSA.  For a foreground screen, the free--free optical depth is
\begin{equation}
\tau_{\rm ff}\simeq 0.08235
\left(\frac{T_e}{10^4\,{\rm K}}\right)^{-1.35}
\left(\frac{\nu}{\rm GHz}\right)^{-2.1}
\left(\frac{{\rm EM}}{{\rm pc\,cm^{-6}}}\right),
\end{equation}
with emission measure ${\rm EM}\equiv\int n_e^2\,dl$ \citep[e.g.,][]{1967ApJ...147..471M}.  Reaching $\tau_{\rm ff}\sim1$ at $\nu\simeq6$--9\,GHz requires ${\rm EM}\sim{\rm few}\times10^{8}$--$10^{9}\ {\rm pc\,cm^{-6}}$ for $T_e\sim10^4$\,K, i.e., very dense ionized gas on sub-parsec scales:
\begin{equation}
n_e \sim 3\times10^4\ 
\left(\frac{{\rm EM}}{10^{9}\,{\rm pc\,cm^{-6}}}\right)^{1/2}
\left(\frac{L}{10^{-3}\,{\rm pc}}\right)^{-1/2}\ {\rm cm^{-3}}.
\end{equation}
Such conditions are not implausible in clumpy torus/disk-wind material and have been inferred in a handful of nearby AGN where FFA produces strong spectral curvature or jet/counter-jet asymmetries \citep[e.g.,][]{2000ApJ...530..233W,2001PASJ...53..169K}.

However, the persistent low-frequency component imposes a stringent covering-factor constraint.  Any uniform FFA screen that makes the flare optically thick near several GHz would render the steady component opaque at $\lesssim1$\,GHz, contrary to the observed long-lived low-frequency spectrum.  If FFA contributes at all, it must therefore be patchy and preferentially cover the newly emerged compact component.  This is a useful prediction rather than a nuisance: simultaneous broadband spectra can test for the exponential cutoff expected from FFA, while polarization and rotation-measure monitoring can probe whether dense ionized clumps move into (or out of) the line of sight as the flare evolves.

\subsection{What does the flare evolution imply?} \label{sec:4.2}

The flare turnover evolves non-monotonically: $\nu_{\rm p,flare}$ rises from $\sim6.3$\,GHz (Epoch~I) to $\sim8.6$\,GHz (Epoch~II) and then drops to $\sim5.8$\,GHz (Epoch~III), while $S_{\rm p,flare}$ rises and then declines.  In the simplest adiabatically expanding synchrotron-blob picture (van der Laan-type evolution), $\nu_{\rm p}$ is expected to drift steadily downward with time.  The observed rise therefore points to continued energization during 2022: either additional injection of non-thermal particles, an increase in magnetic field strength/ordering, or the superposition of multiple unresolved sub-flares.  By Epoch~III the trends reverse ($R_{\rm eq}$ increases while $B_{\rm eq}$ decreases), consistent with the flare entering a decay phase dominated by expansion and declining field strength.

The radio peak occurs months after the early-2022 optical and X-ray activity; the delay to the late-2022 maximum is of order $\sim260\pm40$\,days.  We do not interpret this as a single ballistic travel time.  In self-absorbed synchrotron sources, apparent delays naturally arise because the emission becomes transparent at a given frequency only after the source has expanded or the particle distribution has evolved.  The compactness implied by the turnover ($R_{\rm eq}\sim10^{-3}$\,pc) also means that a months-long flare cannot be explained by a single impulsive ejection expanding rapidly and then cooling in isolation; some combination of confinement, slow expansion, or sustained/repeated energy injection is required (Appendix~\ref{app:phys}).

\subsection{Compact scenarios consistent with the constraints}

The data therefore favor compact, continuously (or repeatedly) energized scenarios.  Two models warrant consideration:

\noindent\textit{(A) Jet-base plasmon / mini-jet flare.}  A compact knot is injected near the jet base during (or shortly after) the early-2022 high-energy episode.  The months-long growth phase then reflects ongoing particle acceleration and magnetic-flux buildup in an initially opaque core, while the later decline corresponds to expansion and adiabatic losses once injection weakens.  A natural observational test is structural evolution: a new VLBI component should eventually separate from the core or produce a measurable frequency-dependent core shift.  Our 1.7-yr monitoring constrains any resolvable motion to $\beta_{\rm app} \equiv v/c \lesssim0.5$--0.8, but a slow ($\lesssim0.2c$) or still-unresolved ejecta remains allowed.  Compact, mm-peaking flares linked to nascent jet components are well documented in III~Zw~2 \citep{1999ApJ...514L..17F, 2005A&A...435..497B, 2022arXiv221213735W} and in weakly jetted nuclei \citep[e.g., Sgr~A*,][]{2006ApJ...650..189Y, 2008ApJ...682..361Y}.

\noindent\textit{(B) Outflow--CNM shock.}  A fast (sub-relativistic) outflow launched during the high-energy episode dissipates as it impacts the surrounding gas.  The steep CNM cusp inferred above provides a natural way to keep the dissipation compact: high ambient pressure near the nucleus can confine the shock and boost its radiative efficiency at high frequency, while a declining density with radius allows the turnover to drift downward once the outflow propagates outward.  This picture is also compatible with the largely stationary VLBI centroid.  Self-absorbed peaks with subsequent downward frequency drift are generic signatures of shock-powered radio transients, from radio supernovae to non-jetted tidal disruption events \citep[e.g.,][]{1998ApJ...499..810C, 2016ApJ...819L..25A, 2021ApJ...919..127C, 2022ApJ...927...74M}.

\subsection{Radio--X-ray/optical connection and implications for a putative pre-coalescence SMBHB}

The delayed radio flare is best viewed as a kinetic echo of the early-2022 high-energy episode: it measures when, and where, a fraction of the injected energy is processed into relativistic electrons and magnetic field.  In an SMBHB framework, strong time-dependent perturbations (e.g., stream impacts, shocks at a circumbinary cavity edge, or mini-disk interactions) can plausibly trigger episodic outflows.  In a single-SMBH framework, an extreme inner-disk/coronal disturbance can have the same results.  Either way, the radio data add an important piece that optical/X-ray light curves alone cannot supply: they demonstrate that the event couples to a compact, synchrotron-emitting outflow channel and that this channel is shaped by a dense, structured CNM on $\ll$\,pc scales.

The observed sequence, from a dramatic high-energy disturbance to a compact high-frequency-peaking radio flare a few months later, has close phenomenological analogues in changing-look AGN \citep{2023NatAs...7.1282R, 2025ApJ...979L...2M} and in outflow-powered tidal disruption events \citep[e.g.,][]{2016ApJ...819L..25A, 2021ApJ...919..127C}.  J1430+2303 shows that similar radiative--kinetic coupling can occur in a radio-quiet nucleus and remain confined to sub-parsec scales. In this broader sense, J1430+2303 may represent a nominally radio-quiet nucleus caught during a transient radio-bright phase (An 2026, under review). If future monitoring reveals recurrent flares tied to repeat optical/X-ray activity, their cadence and spectral evolution will offer a purely electromagnetic way to track repeated energy injection in a system that has been discussed as an SMBHB candidate.  Regardless of the underlying driver, the combination of VLBI localization and broadband turnover evolution provides a concrete observational template for identifying and interpreting rapidly changing galactic nuclei in the era of low-frequency gravitational-wave astronomy.

\section{Summary}
\label{sec:summary}

We presented multi-epoch VLBI imaging and time-resolved broadband radio spectra of J1430+2303, focusing on the emergence and evolution of a compact, self-absorbed nuclear flare that is delayed with respect to the early-2022 optical/X-ray activity.  The main results are:

\begin{enumerate}
    \item Across 4.7--22.2\,GHz and over 2022--2024, the VLBI emission is consistently dominated by a single unresolved core with $T_{\rm B}\gtrsim10^{7}$\,K.  No secondary component is detected; over $\simeq1.7$\,yr this constrains any resolvable proper motion to $\lesssim0.5$--$0.8\,c$ (and $\lesssim0.16\,c$ if interpreted as unresolved expansion).
    \item A near-epoch VLBA/VLA comparison at 15\,GHz recovers $\sim76\%$ of the connected-array flux density, directly tying the high-frequency brightening to the milliarcsecond core.  Subsequent VLBA monitoring at 15.2\,GHz reveals renewed brightening between 2023 September and 2024 February while the source remains unresolved, pointing to episodic energy injection inside the central parsec.
    \item The connected-array spectra require a two-component model: a persistent low-frequency SSA component with $\nu_{\rm p,steady}\simeq0.74$\,GHz and a time-variable flare component.  The flare turnover evolves from $\sim6$\,GHz (early 2022) to $\sim9$\,GHz (late 2022) and then back to $\sim6$\,GHz (Spring 2023).
    \item Equipartition scalings across these two components reveal a steep inner circumnuclear density cusp ($n \propto R^{-1.7}$) mapping from the steady emission zone ($R \sim 9 \times 10^{-3}$~pc) down to the deeply embedded flare environment ($R \sim 5 \times 10^{-4}$~pc).  The inferred magnetic energy of the Epoch~II flare is $E_B\sim{\rm few}\times10^{44}$\,erg, implying an internal-energy scale $\sim{\rm few}\times10^{45}$\,erg for fiducial $\epsilon_B\sim0.1$.
    \item The months-long delay relative to the high-energy trigger and the sustained, compact nature of the flare argue against a single, freely expanding impulsive ejection as the sole explanation.  The data are instead naturally accommodated by compact scenarios that involve confinement or continued energization: a jet-base plasmon that remains unresolved for years, or an outflow that dissipates through shocks in a dense, structured CNM.  A uniform free--free absorber is disfavored by the survival of the steady low-frequency component; if external absorption contributes, it must be patchy and partially covering.  Continued multi-frequency VLBI and polarization monitoring (core shift, rotation-measure variability, and the possible appearance of a slow knot) provide direct tests.
\end{enumerate}

\begin{acknowledgements}
T.A. thanks for the Shanghai Oriental Talent Project. Y.Z. is sponsored by Shanghai Sailing Program (22YF1456100). 
A.W. supported by the National Natural Science Foundation of China under Grant Number 12503018 and the China Postdoctoral Science Foundation under Grant Number of 2025M773197 and 2025T180874.
We thank the TACs and schedulers of the EVN and NRAO for approving the ToO/DDT observations and quickly scheduling the observations. 
The European VLBI Network (EVN) is a joint facility of independent European, African, Asian, and North American radio astronomy institutes. Scientific results from data presented in this publication are derived from the following EVN project code(s): RS005, RA006. 
The data presented in this paper are based on observations
made with the Karl G. Jansky Very Large Array from the program VLA/21B-375 and VLA/22B-106, 
and the Giant Metrewave Radio Telescope from the project ddtC224. 
We thank the help from the staff of the VLA, VLBA and GMRT that made these observations possible. 
The National Radio Astronomy Observatory is a facility of the National Science Foundation operated under cooperative agreement by Associated Universities, Inc. Scientific results from data presented in this publication are derived from the following VLBA project codes: BA154, BA157, BA166.
This work used resources from the China SKA Regional Centre prototype \citep{2019NatAs...3.1030A, 2022SCPMA..6529501A}.
GMRT is run by the National Centre for Radio Astrophysics of the Tata Institute of Fundamental Research. 
\end{acknowledgements}

%
  \bibliographystyle{aa} 
  \bibliography{J1430radio.bib} 
%

\clearpage

\begin{table*}
 	\caption{The observation information. \label{tab:VLBIobs}}
 	\centering
 	\begin{tabular}{ccccccccc}
 		\hline\hline
Code   &    date    & $\nu$ & bandwidth & $\tau_{\rm obs}$ &            Antenna   & Ref.   \\
       &   Y-M-D    & (GHz) &   (MHz)   &   (hour)         &                      &        \\ \hline
RA005   & 2022-02-27 & 1.7  &    256    &       5.5        &     EVN (14 telescopes)       & 1 \\
BA154a  & 2022-03-02 & 1.6  &    512    &       4.5        & SC,FD,LA,BR,HN,NL,PT,KP,OV,MK & 1 \\ 
BA154b  & 2022-03-02 & 4.9  &    512    &       4.5        & SC,FD,LA,BR,HN,NL,PT,KP,OV,MK & 1 \\
BA157b  & 2022-04-26 & 4.7  &    512    &       4.0        & BR,FD,HN,KP,LA,MK,NL,OV,PT,SC & 2 \\
BA157b  & 2022-04-26 & 7.6  &    512    &       4.0        & BR,FD,HN,KP,LA,MK,NL,OV,SC    & 2 \\
BA157a  & 2022-05-08 &15.2  &   1024    &       6.0        & BR,FD,KP,MK,NL,OV,LA          & 2 \\
BA157a1 & 2022-05-31 &15.2  &   1024    &       6.0        & BR,FD,KP,MK,NL,OV,PT,SC       & 2 \\ 
RA006   & 2023-02-16 &22.2  &   512    &       6.0        & EVN (10 telescopes)           & 2 \\ 
BA166a   & 2023-09-01 &15.2  &   1024    &       6.0        & BR,FD,HN,KP,LA,MK,NL,OV,PT           & 2 \\ 
BA166b   & 2023-12-02 &15.2  &   1024    &       6.0        & BR,FD,HN,KP,LA,MK,NL,OV,PT,SC           & 2 \\ 
BA166c   & 2024-02-18 &15.2  &   1024    &       6.0        & BR,FD,HN,KP,LA,MK,NL,OV,PT           & 2 \\ \hline
\end{tabular}\\
 	Notes: Col.~1 -- Project code;  Col.~2 -- Observation date; Col.~3 -- Central frequency; Col.~4 --  The bandwidth used for the imaging; Col.~5 -- The observation time; Col.~6 -- The telescopes participated in the corresponding observations. The complete VLBA antennas and their name abbreviations are BR (Brewster), FD (Fort Davis), HN (Hancock), KP (Kitt Peak), LA (Los Alamos), MK (Mauna Kea), NL (North Liberty), OV (Owens Valley), PT (Pie Town) and SC (Saint Croix). The EVN telescopes that participated in the observation RS005 are referred to \citet{2022A&A...663A.139A}, and the EVN telescopes that participated in RA006 are described in Appendix \ref{app:obs}. References: 1 -- \citet{2022A&A...663A.139A}; 2 -- the present paper. 
\end{table*}

\begin{table*}[h]
	\caption{Parameters of J1430+2303 obtained from \textsc{Difmap} and \textsc{MCMC} model fitting. \label{tab:mod}}
	\centering
    \setlength{\tabcolsep}{3pt}
	\begin{tabular}{lcccccclccccc}
		\hline\hline
Code   & $\nu$ & Beam                      & rms noise       & $S_{\rm peak}$   & $S_{\rm total}^{\rm difmap}$ & $S_{\rm total}^{\rm mcmc}$ & $\theta^{\rm difmap}$ & $\theta^{\rm mcmc}$    & $T_\mathrm{B}$        &  Ref. \\     
       & (GHz) & (mas$\times$mas, $\degr$) & (mJy\,b$^{-1}$) & (mJy\,b$^{-1}$)  &       (mJy)                  &    (mJy)                   &   (mas)               &   (mas)                & ($\times$10$^{7}$K)   &       \\ 
       (1) & (2) & (3) & (4) & (5) & (6) & (7) & (8) & (9) & (10) & (11) \\ \hline
ra005  &  1.7  & $7.6\times2.9$, $13.5$    & 0.015           &   0.285$\pm$0.022  &  0.328$\pm$0.030               &0.325$\pm$0.018             & $\le$1.02              &1.19$\pm$0.33          & 10.5$\pm$7.9           & 1 \\         
ba154a &  1.6  &$11.7\times5.1$, $1.6$     & 0.018           &   0.220$\pm$0.022  &  0.222$\pm$0.029               &0.227$\pm$0.017             & $\le$2.17              &$\le$0.98              & $\ge$12.2              & 1 \\         
ba154b &  4.9  & $3.9\times1.7$, $-5.4$    & 0.008           &   0.238$\pm$0.014  &  0.243$\pm$0.017               &0.242$\pm$0.008             & $\le$0.45              &$\le$0.31              & $\ge$13.8              & 1 \\         \hline
ba157b &  4.7  & $3.9\times1.6$, $1.2$     & 0.017           &   0.214$\pm$0.020  &  0.214$\pm$0.026               &0.216$\pm$0.013             & $\le$0.67             &$\le$0.19              & $\ge$35.7              & 2 \\         
ba157b &  6.2  & $3.0\times1.3$, $0.6$     & 0.012           &   0.191$\pm$0.015  &  0.192$\pm$0.020               &0.188$\pm$0.010             & $\le$0.48             &$\le$0.25              & $\ge$10.3              & 2 \\         
ba157b &  7.6  & $2.3\times0.9$, $0.0$     & 0.020           &   0.176$\pm$0.022  &  0.180$\pm$0.030               &0.181$\pm$0.016             & $\le$0.47             &$\le$0.21              & $\ge$9.4               & 2 \\         
ba157a & 15.2  & $1.6\times0.7$, $-15.5$   & 0.034           &   0.186$\pm$0.036  &  0.200$\pm$0.051               &0.209$\pm$0.043             & $\le$0.45             &$\le$0.34              & $\ge$1.0               & 2 \\         
ba157a1& 15.2  & $1.2\times0.5$, $-4.0$    & 0.028           &   0.191$\pm$0.030  &  0.208$\pm$0.043               &0.208$\pm$0.031             & $\le$0.29             &$\le$0.22              & $\ge$2.5               & 2 \\
RA006  & 22.2  & $0.8\times0.4$, $-2.8$    & 0.072           &   0.353$\pm$0.072  &  0.365$\pm$0.072               &0.338$\pm$0.081             & $\le$0.32             & ...                   & $\ge$1.0               & 2 \\ 
ba166a& 15.2  & $1.5\times0.6$, $-18.3$    & 0.030           &   0.154$\pm$0.031  &  0.150$\pm$0.042               &0.154$\pm$0.032             & $\le$0.41             &$\le$0.38              & $\ge$0.5               & 2 \\
ba166b& 15.2  & $1.2\times0.5$, $-4.0$    & 0.030           &   0.228$\pm$0.032  & 0.227$\pm$0.044               &0.238$^{+0.035}_{-0.034}$             & $\le$0.27             &$\le$0.29              & $\ge$1.8               & 2 \\
ba166c& 15.2  & $1.2\times0.5$, $-4.0$    & 0.027           &   0.299$\pm$0.031  &  0.297$\pm$0.041               &0.305$^{+0.029}_{-0.028}$             & $\le$0.31             &$\le$0.27              & $\ge$1.8               & 2 \\
\hline  
	\end{tabular}\\
    Col.~1 -- project code; Col.~2 -- central frequency; Col.~3 -- restoring beam of the image; Col.~4 -- \textit{rms} noise of the image; Col.~5 -- peak flux density of the component; Col.~6 and 7 -- total flux density of the component derived from \textsc{Difmap} and \textsc{MCMC}, respectively; Col.~8 and 9 -- upper limit of the source size estimated from \textsc{Difmap} and \textsc{MCMC}, respectively;  Col.~10 -- brightness temperature of the VLBI component; Col.~11 -- reference of the data. 
\end{table*}

\begin{table*}[h]
	\caption{Summary of connected-array radio observations of J1430+2303.}
	\label{tab:vla_gmrt}
	\centering
    \setlength{\tabcolsep}{3pt}
	\begin{tabular}{lcccccl}
    \hline \hline
Code         & date        & $\nu$  & Beam                           & rms noise             & $S_{\rm peak}$   & $S_{\rm total}$  \\ 
 & & (GHz) & (\arcsec, \arcsec, $\degr$) & ($\mu$Jy beam$^{-1}$) & ($\mu$Jy beam$^{-1}$) & ($\mu$Jy)\\ \hline
VLA/21B-375  & 2022-04-06  & 8.5    & 0.011 $\times$ 0.028, 78       & 13              & 326              & 352 $\pm$ 6      \\
             &             & 9.5    & 0.074 $\times$ 0.027, 78       & 12              & 316              & 332 $\pm$ 2      \\
             & 2022-04-13  & 5.0    & 0.57 $\times$ 0.32, 69         & 17              & 397              & 418 $\pm$ 3      \\
             &             & 6.0    & 0.48 $\times$ 0.26, 70         & 18              & 403              & 412 $\pm$ 10     \\
             & 2022-05-04  & 2.5    & 0.73 $\times$ 0.61, $-$65      & 21              & 457              & 473 $\pm$ 7      \\
             &             & 3.5    & 0.55 $\times$ 0.47, $-$66      & 16              & 458              & 496 $\pm$ 9      \\
             & 2022-05-06  & 12.7   & 0.24 $\times$ 0.12, 71         & 17              & 258              & 283 $\pm$ 7      \\
             &             & 14.3   & 0.21 $\times$ 0.12, 68         & 15              & 226              & 219 $\pm$ 13     \\
             &             & 15.7   & 0.19 $\times$ 0.11, 68         & 16              & 205              & 226 $\pm$ 10     \\
             &             & 17.3   & 0.18 $\times$ 0.10, 70         & 18              & 160              & 150 $\pm$ 6      \\
VLA/22B-106  & 2022-12-26  & 4.5    & 4.03 $\times$ 3.69, $-$3       & 35              & 501              & 522 $\pm$ 6      \\
             &             & 6.5    & 2.89 $\times$ 2.59, 7          & 30              & 478              & 476 $\pm$ 8      \\
             &             & 7.5    & 2.55 $\times$ 2.27, 12         & 24              & 463              & 486 $\pm$ 13     \\
             &             & 2.5    & 2.58 $\times$ 1.86, 161        & 69              & 640              & 720 $\pm$ 10     \\
             &             & 3.5    & 5.07 $\times$ 4.61, 4.89       & 46              & 590              & 590 $\pm$ 10     \\
             & 2022-12-27  & 12.8   & 0.69 $\times$ 0.43, 111        & 21              & 394              & 432 $\pm$ 10     \\
             &             & 14.3   & 1.96 $\times$ 1.26, $-$63      & 22              & 431              & 435 $\pm$ 8      \\
             &             & 15.9   & 1.80 $\times$ 1.15, $-$65      & 22              & 421              & 416 $\pm$ 16     \\
             &             & 17.4   & 1.69 $\times$ 1.10, $-$61      & 26              & 433              & 461 $\pm$ 23     \\
             & 2023-03-07  & 15.1   & 0.15 $\times$ 0.08, 143        & 12              & 296              & 322 $\pm$ 7      \\
             &             & 10.0   & 0.55 $\times$ 0.49, 5          & 14              & 398              & 410 $\pm$ 5      \\
             & 2023-03-09  & 6.0    & 0.35 $\times$ 0.19, 131        & 19              & 449              & 481 $\pm$ 11     \\ 
             & 2023-04-26  & 13.5   & 0.45 $\times$ 0.40, $-$56      & 16              & 212              & 226 $\pm$ 7      \\
             &             & 16.6   & 0.38 $\times$ 0.33, $-$64      & 19              & 161              & 202 $\pm$ 19     \\
             & 2023-04-26  & 9.0    & 0.62 $\times$ 0.57, $-$6       & 18              & 389              & 421 $\pm$ 6      \\
             &             & 11.0   & 0.52 $\times$ 0.47, 8          & 19              & 331              & 355 $\pm$ 6      \\
             &             & 5.0    & 0.40 $\times$ 0.24, 30         & 22              & 482              & 532 $\pm$ 7      \\
             &             & 7.0    & 0.29 $\times$ 0.17, 147        & 21              & 388              & 426 $\pm$ 13     \\
GMRT/ddtC244 & 2022-03-09  & 1.3    & 1.30 $\times$ 0.91, 106        & 37              & 1002             & 1185 $\pm$ 23      \\
             & 2022-03-15  & 0.75   & 1.49 $\times$ 0.70, 159        & 46              & 1115             & 1292 $\pm$ 45      \\
             & 2022-12-26  & 1.3    & 2.04 $\times$ 1.92, 139        & 23              & 838              & 880 $\pm$ 12       \\
             & 2022-12-27  & 0.75   & 2.05 $\times$ 1.24, 88         & 53              & 1026             & 1198 $\pm$ 49      \\
VLASS        & 2017-11-17  & 3.0    & 2.51 $\times$ 2.20, 52         & 116             & 430              & 430 $\pm$ 116       \\
             & 2020-07-16  & 3.0    & 2.38 $\times$ 2.10, $-$11      & 130             & 441              & 441 $\pm$ 130       \\
             & 2023-01-23  & 3.0    & 2.51 $\times$ 2.20, 52         & 115             & 581              & 581 $\pm$ 115       \\
ASKAP/RACS   & 2022-01-05  & 1.7    & 16.90 $\times$ 8.90, 152       & 174             & 683              & 683 $\pm$ 174      \\
             & 2022-07-29  & 1.4    & 21.09 $\times$ 10.60, 2        & 181             & 905              & 905 $\pm$ 181      \\   
             & 2020-10-16  & 0.888  & 25.00 $\times$ 25.00           & 299             & 1031             & 1031 $\pm$ 299     \\
             & 2020-10-17  & 0.888  & 25.00 $\times$ 25.00           & 299             & 1044             & 1044 $\pm$ 299      \\
\hline  
	\end{tabular}\\
\end{table*}

\begin{table}
\centering
\caption{SSA decomposition parameters and derived physical parameters.}
\label{tab:ssa_components}
\begin{tabular}{lcccc}
\hline\hline
Component & $\nu_p$ (GHz) & $F_{\nu,p}$ (mJy) & $R_{\rm eq}$ (pc) & $B_{\rm eq}$ (G)\\
\hline
Steady & $0.740^{+0.090}_{-0.110}$ & $1.221^{+0.120}_{-0.080}$ & $9.14\times10^{-3}$ & 0.102 \\
Flare (Epoch I) & $6.346^{+0.593}_{-0.570}$ & $0.176^{+0.024}_{-0.022}$ & $4.24\times10^{-4}$ & 1.081 \\
Flare (Epoch II) & $8.608^{+0.851}_{-0.744}$ & $0.384^{+0.027}_{-0.028}$ & $4.53\times10^{-4}$ & 1.351 \\
Flare (Epoch III) & $5.831^{+0.646}_{-0.645}$ & $0.253^{+0.030}_{-0.025}$ & $5.49\times10^{-4}$ & 0.956 \\
\hline
\end{tabular}
\end{table}

\begin{appendix}

\section{VLBI Observations and data reduction} \label{app:obs}

\begin{figure*} 
  \centering
  \includegraphics[width=0.32\textwidth]{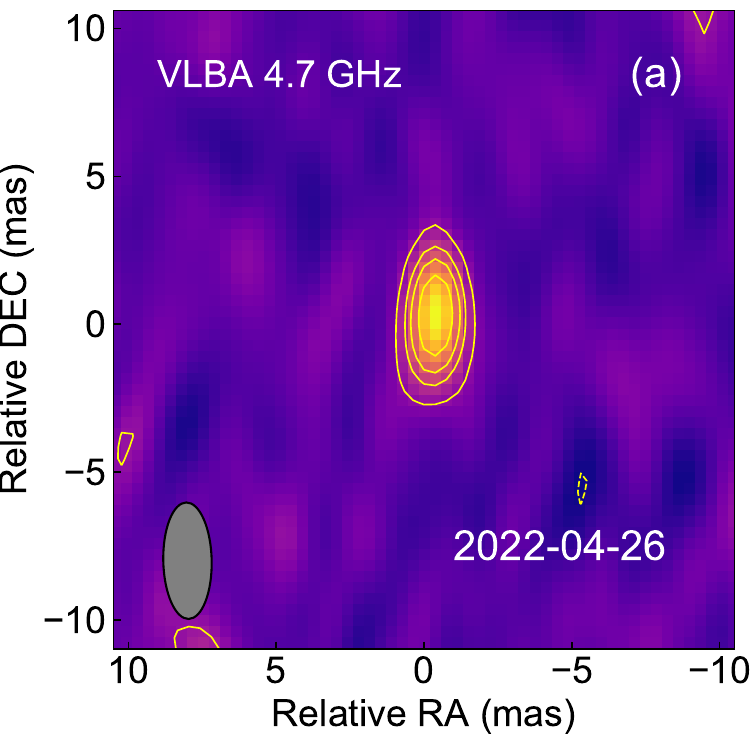}
  \includegraphics[width=0.32\textwidth]{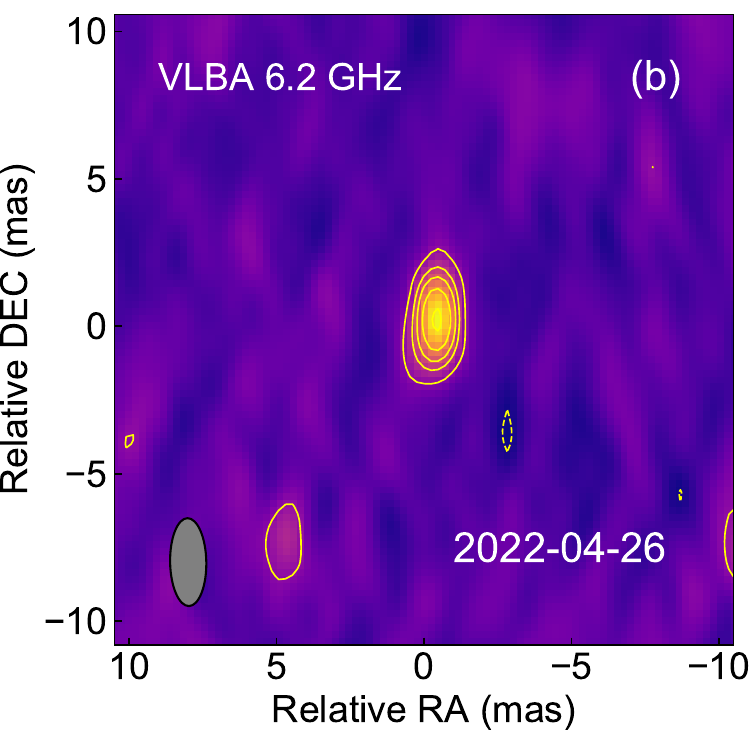}
  \includegraphics[width=0.32\textwidth]{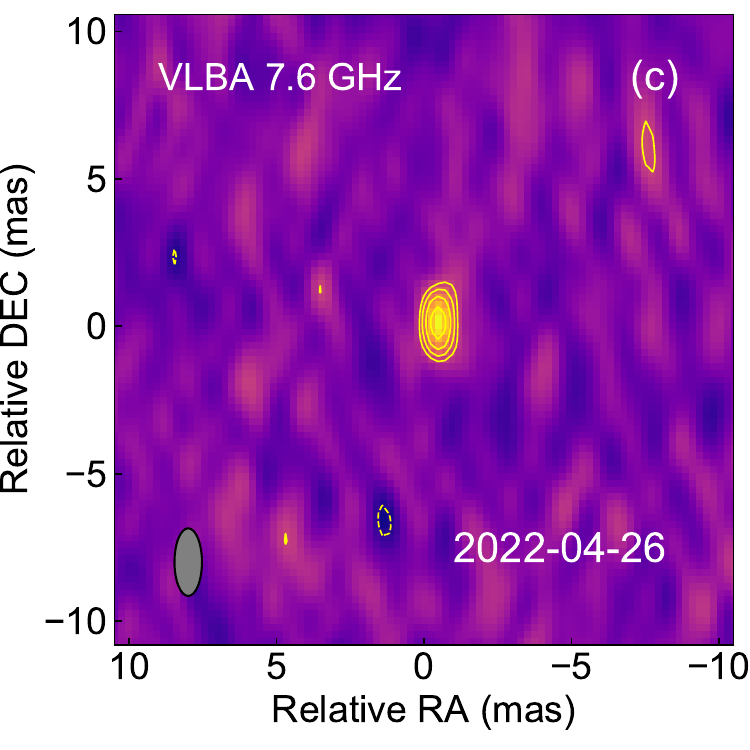} \\
  \includegraphics[width=0.32\textwidth]{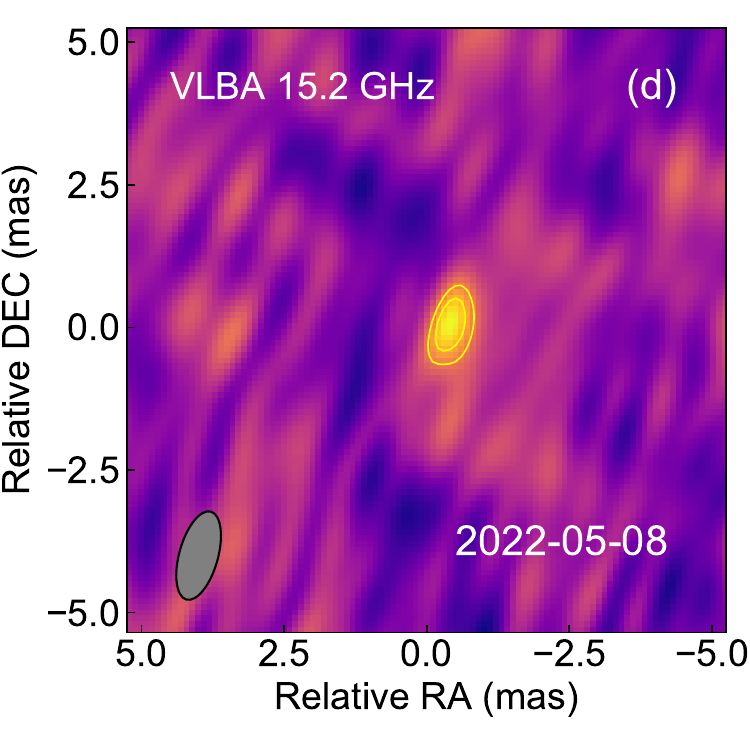}
  \includegraphics[width=0.32\textwidth]{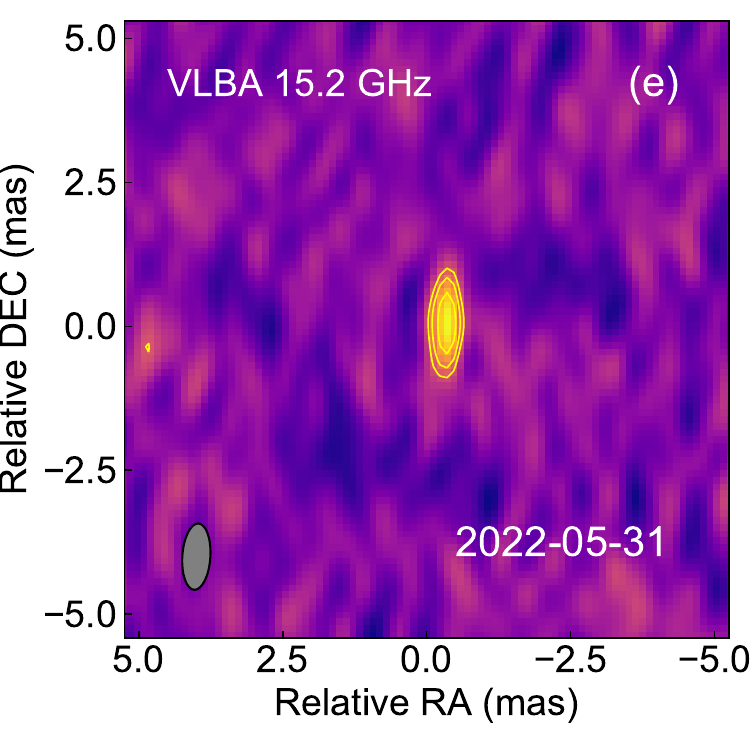}
  \includegraphics[width=0.32\textwidth]{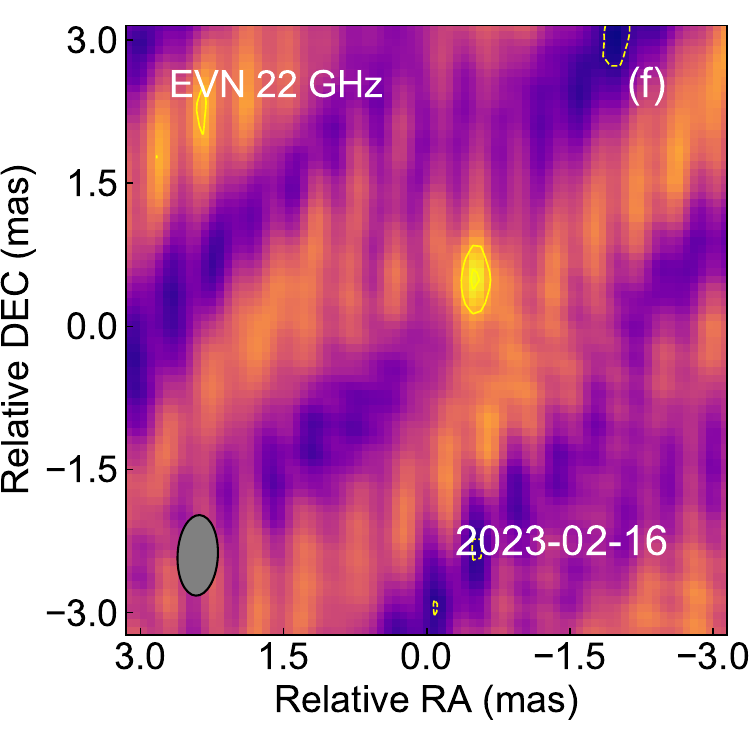} \\
  \includegraphics[width=0.32\textwidth]{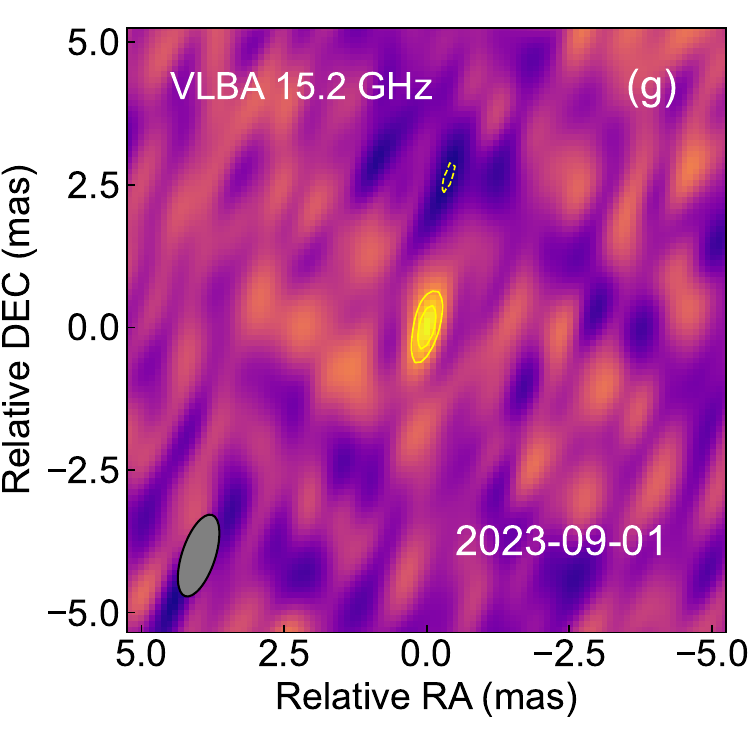}
  \includegraphics[width=0.32\textwidth]{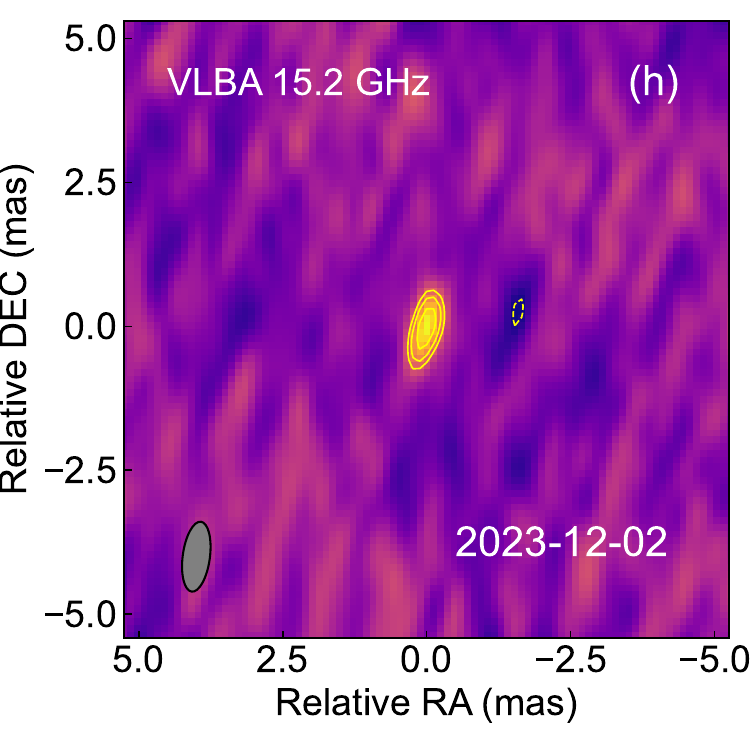}
  \includegraphics[width=0.32\textwidth]{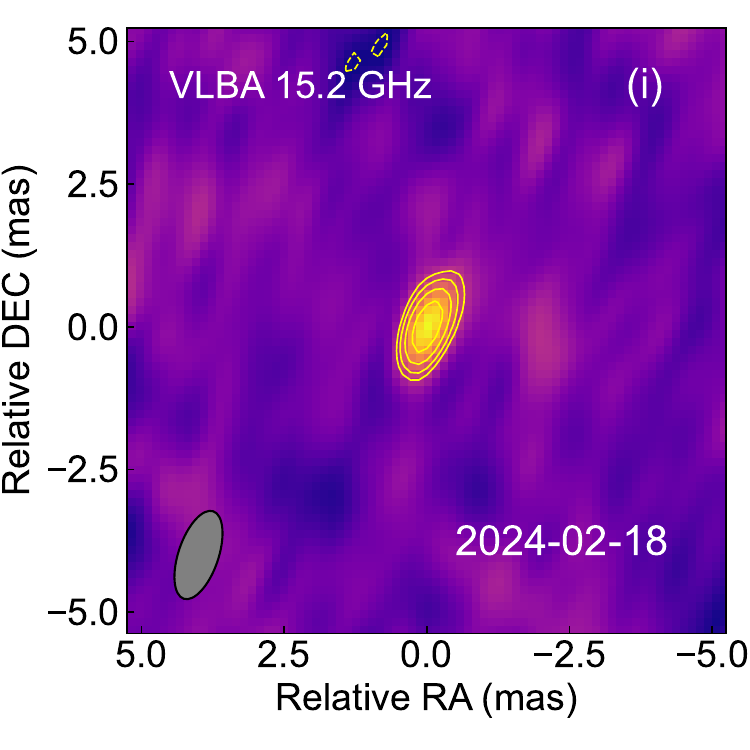} \\
  \caption{Natural-weighted VLBI images of J1430$+$2303. \textit{(a) - (c)}:  VLBA images at 4.7, 6.2 and 7.6 GHz, respectively; \textit{(d) - (e)}: VLBA images at 15.2 GHz observed on 8 and 31 May 2022, respectively; (f) 22-GHz EVN image observed on 16 February 2023. The ellipse in the bottom-left corner is the shape of the restoring beam. The contours start at 3$\sigma$ and increase in a step of $2^{1/2}$. 
  The observation information are listed in Table~\ref{tab:VLBIobs} and the image parameters are presented in Table~\ref{tab:mod}.}
  \label{fig:tar}
\end{figure*}

We obtained three new VLBI projects on J1430+2303 (Table \ref{tab:VLBIobs}): (i) BA157 with the VLBA at C band, using two sub-bands centered at 4.7 and 7.6~GHz to enable in-band spectral-index mapping, and at U band at 15.2~GHz; (ii) BA166 with the VLBA at U band (15.2~GHz); and (iii) RA006 with the EVN at K band (22.2~GHz). All sessions were executed in phase-referencing mode using J1427+2348 (separation $\sim$1$^\circ$) as the phase calibrator. A strong fringe finder (e.g.  3C~345) was observed at the beginning of each run to derive instrumental delays and phases.

In the VLBA/BA157b session we imaged at 4.7 and 7.6~GHz and also formed a 6.2~GHz map by concatenating the two sub-bands; the 6.2~GHz image has the lowest noise owing to the combined bandwidth. 

\smallskip
\noindent\textit{Frequency setups and data rates.} 
The VLBA sessions (BA157 and BA166) recorded a total bandwidth of 1024~MHz with a data rate of 4096~Mbps per antenna. The EVN session (RA006) at 22.2~GHz used e-VLBI with a data rate of 2048~Mbps. For display purposes, the two C-band sub-bands of BA157 were also concatenated into an effective \mbox{$\sim$6.2~GHz} continuum image; however, spectral-index measurements always used the two sub-bands reduced separately.

\smallskip
\noindent\textit{Phase-referencing cycles and on-source time.}
Typical cycles were 5~min at C band, Calibrator (\textit{Cal})~1~min, Target (\textit{Tar})~4~min, a shorter cycle $\sim$90~s at U band (\textit{Cal}~20~s; \textit{Tar}~70~s), and 2~min at K band (\textit{Cal}~1~min; \textit{Tar}~1~min). The net on-source times were approximately 2.5~h (C band), 3.0--3.2~h (U band, for BA157/BA166 individually), and 2.7~h (K band). Participating antennas followed standard VLBA and EVN availability for these bands; any station-level data-loss incidents are summarized in Table~\ref{tab:VLBIobs}. Final image noise levels and restoring beams are reported in Table~\ref{tab:mod} and in the figure captions.

Raw data were correlated with \textsc{DiFX} correlator  \citep{2011PASP..123..275D} at the VLBA Data Correlation Center in Socorro, USA, with an integration time of 2~s and with the \textsc{SFXC} at Joint Institute for VLBI European Research Infrastructure Consortium (JIVE) in Dwingeloo, the Netherlands for the EVN (integration time 1~s). The correlated visibilities were then transferred to the China SKA Regional Centre \citep{2019NatAs...3.1030A} and processed with our automated VLBI pipeline\footnote{\url{https://github.com/SHAO-SKA/vlbi-pipeline}}, a \textsc{Python}+\textsc{Parseltongue} interface to \textsc{AIPS} \citep{2003ASSL..285..109G} following the standard \textsc{AIPS} Cookbook \footnote{\url{http://www.aips.nrao.edu/cook.html}}.

For all sessions, the calibration sequence was: (1) a~priori amplitude calibration with \textsc{apcal}, including weather-based opacity and antenna-related gain curve corrections; (2) Earth-orientation and atmospheric updates via \textsc{clcor}; (3) instrumental delay/phase from a bright fringe finder (3C~345); (4) global fringe fitting on the phase reference calibrator J1427+2348 with \textsc{fring}; and (5) antenna bandpass calibration from 3C~345 using \textsc{bpass}. For the wide C~band (BA157b), we first ran the full-band pipeline to form a $\sim$6.2\,GHz continuum image and then processed the two sub-bands (centered at 4.7/7.6\,GHz) separately.

Data quality notes: PT station of the VLBA at 7.6\,GHz was removed from calibration due to a synthesizer failure. In RA006, Jb and Nt did not contribute due to Fila10G and DBBC issues, and several Kerean VLBI Network stations ceased participation around February~16 02:07~UT; the final array used for imaging comprised Ef, Mc, O6, Ur, Ys, Mh, and Hh.

The calibrator J1427+2348 was exported as averaged single-source data (128\,MHz channels; 2\,s) to \textsc{Difmap} \citep{1994BAAS...26..987S} for iterative CLEAN and self-calibration (phase-only to 5$\sigma$ residuals, then phase+amplitude self-calibration with solution intervals reduced  to $\sim$4\,min). The resulting CLEAN-component model was re-imported into \textsc{AIPS} for a final fringe fitting, mitigating source structure-related phase errors in the phase reference. All solutions were then applied to the target \tar, which was imaged in \textsc{Difmap}. Given the low S/N of \tar, no self-calibration was attempted.

\begin{figure*}
    \centering
    \includegraphics[height=8cm]{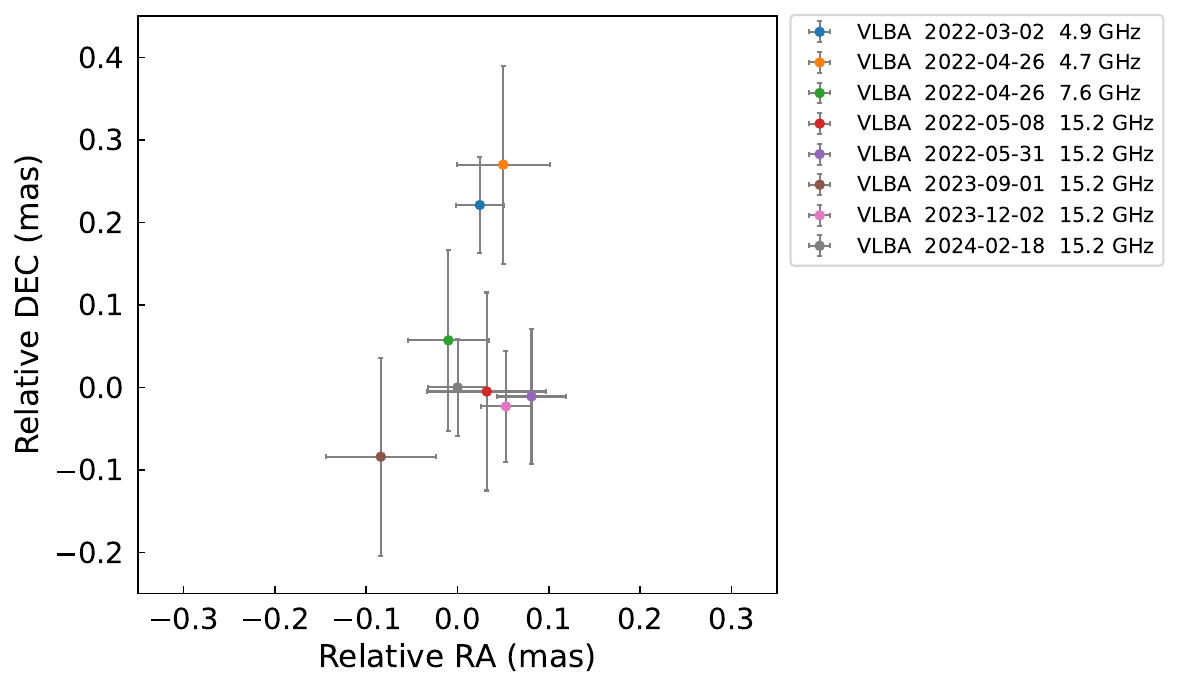}
    \caption{Relative astrometric positions of the VLBI core across epochs. The origin (0,0) is the core position measured at the last epoch (14$^h$30$^m$16.0409$^s$, +23$^d$03$^m$44.53875$^s$). Error bars show the 1$\sigma$ uncertainties from the \textsc{Difmap} visibility-domain model fits. Measurements from different frequencies within the same epoch are plotted separately.}
    \label{fig:positions}
\end{figure*}

\section{MCMC fitting of VLBI core parameters} \label{app:mcmc}

We model the \tar\ visibilities with a single circular Gaussian in the $(u,v)$ domain. The complex model at baseline $(u,v)$ is
\begin{equation}
    \mathcal{V}_\mathrm{m} = f_\nu e^{-2\pi^2 \left(\frac{\theta}{\sqrt{8\ln 2}}\right)^2\left(u^2+v^2\right)-2\pi i (u\rho + v\delta)}
\end{equation}
where $i=\sqrt{-1}$, $f_\nu$ is the total flux density, $\theta$ the FWHM size, and $(\rho, \delta)$ are offsets relative to the phase center. A simple Gaussian likelihood model was used, namely, 
\begin{equation}
\begin{array}{ll}
    \ln\mathcal{L}(x) = & -\frac{1}{2} \sum_{i=0}^{N}w_i\left[\left(\mathcal{V}_{R,\mathrm{m}}(u_i,v_i,x)-\mathcal{V}_{R,i}\right)^2\right.  \\
    & +\left.\left(\mathcal{V}_{I,\mathrm{m}}(u_i,v_i,x)-\mathcal{V}_{I,i}\right)^2\right] 
\end{array}
\end{equation}
where $\mathcal{V}_{R,i}$ and $\mathcal{V}_{I,i}$ are the real and imaginary parts of the $i$-th visibility measurement, respectively, and $w_i$ is the weight of the data and corresponds to the reciprocal of the square of the associated uncertainty. $\mathcal{V}_{R,\mathrm{m}}(u_i,v_i,x)$ and $\mathcal{V}_{I,\mathrm{m}}(u_i,v_i,x)$ are the real and imaginary parts of the model source visibility evaluated at the point $(u_i,v_i)$. The parameters $x$ is a function $x=(f_\nu,\theta,\rho,\delta)$, where $\theta$ is the full width at half maximum (FWHM). 

We sample the posterior with \textsc{emcee} \citep{2013PASP..125..306F} using 8--16 walkers and $10^4$ steps (discarding the first half as burn-in). Priors are uniform and conservative: $0<f_\nu<5$\,mJy, $0<\theta<2$\,mas, and $|\rho|,|\delta|<2$\,mas. Chains are initialized near the \textsc{Difmap} point-source solution with small Gaussian jitter. Convergence is verified by integrated autocorrelation times and by consistency of split-chain posteriors.

\begin{figure}
    \centering
    \includegraphics[height=8cm]{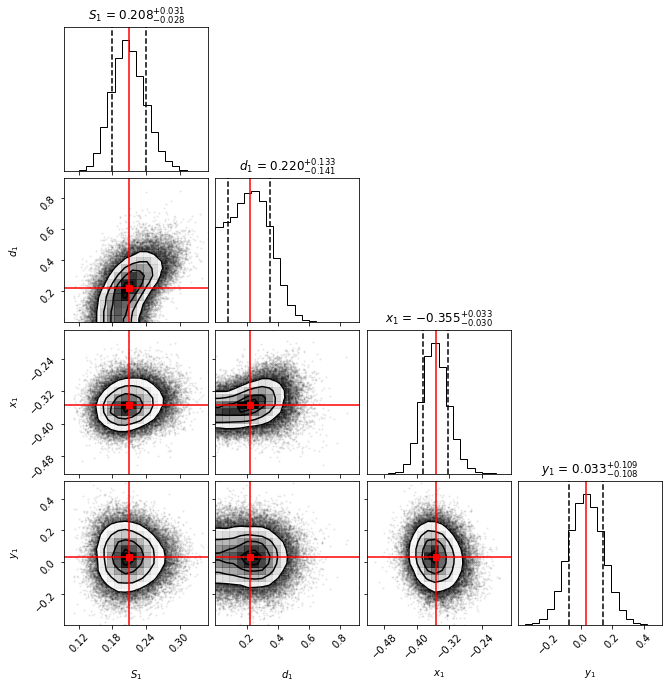}
    \caption{The Corner plot shows all the one and two dimensional projections of the posterior probability distributions of the total flux density ($\rm S_{1}$) in mJy, source size ($\rm d_{1}$) in mas, relative RA ($\rm x_{1}$) in mas, and relative DEC ($\rm y_{1}$) in mas from the 15.2 GHz data (ba157a1; 2022 May 31). The red solid line indicates the best fit and the black dashed line indicates the confidence interval of $1\sigma$.}
    \label{fig:mcmc}
\end{figure}

We report posterior \emph{medians} with 68\% credible intervals ($1\sigma$) (Fig. \ref{fig:mcmc}). For unresolved cases we quote a 95\% upper limit on $\theta$ obtained from the marginalized cumulative distribution function. Flux densities from MCMC agree with image-plane \textsc{Difmap} values within uncertainties. The visibility fit provides more robust size limits at low S/N and naturally accounts for phase-referencing offsets via $(\rho, \delta)$. 

Quality controls include: (i) repeating the fit with natural vs.\ Briggs weighting to check stability of $\theta$; (ii) refitting with outlier-resistant weights (down-weighting $|V_i|>5\sigma$) to verify insensitivity to residual RFI; and (iii) injecting fake point sources into the real $(u,v)$ coverage to confirm correct upper-limit coverage. The adopted parameters are listed in Table~\ref{tab:mod}. Figure~\ref{fig:mcmc} shows a representative corner plot of the visibility-domain MCMC posteriors for the 2022 May 31 epoch (VLBA BA157a1), when the high-frequency synchrotron self-absorption component is apparent. The complete set of corner plots for all bands and epochs is available from the corresponding author on reasonable request.

\section{Physical scalings and order-of-magnitude parameter estimates}\label{app:phys}

This Appendix collects the order-of-magnitude conversions used in the main text and clarifies which inferences rely directly on the SSA turnovers. Throughout we adopt a fiducial expansion speed $v=0.2c$, filling factor $f=0.5$, and equipartition fractions $\epsilon_e=\epsilon_B=0.1$ unless stated otherwise.

\subsection{Brightness Temperature}
The intrinsic brightness temperature of the radio core is calculated as
\begin{equation}
    T_{\rm B} = 1.22 \times 10^{12} (1+z) \left( \frac{S_\nu}{\rm Jy} \right) \left( \frac{\nu}{\rm GHz} \right)^{-2} \left( \frac{\theta}{\rm mas} \right)^{-2} {\rm ~ K},
    \label{eq:TB}
\end{equation}
where $S_\nu$ is the integrated flux density, $\nu$ is the observing frequency, and $\theta$ is the FWHM of the circular Gaussian fitted to the visibilities. The factor of $(1+z)$ corrects for cosmological redshift to yield the intrinsic (source-frame) temperature. 

Because the core remains strictly unresolved across the high-frequency VLBA observations, the true emission region is intrinsically smaller than the synthesized beam. Consequently, the MCMC-derived structural upper limits on $\theta$ dictate strict lower limits on the brightness temperature ($T_{\rm B} \propto \theta^{-2}$). For instance, applying the rigorous MCMC 95\% upper limit for the 15.2~GHz detection in early 2024 ($S_\nu \approx 0.305$~mJy, $\theta < 0.22$~mas), Equation~\ref{eq:TB} implies $T_{\rm B} \gtrsim 3.6 \times 10^7$~K. At lower frequencies where the size limits are more constraining relative to the flux (e.g., at 5~GHz), these lower limits securely exceed $\sim 10^8$~K. These values are highly characteristic of a compact, non-thermal AGN core and strongly disfavors thermal origins (which are restricted to $T_{\rm B} \lesssim 10^5$~K).

\subsection{Steady+flare SSA fitting}\label{app:ssa}
The central idea is to constrain a \emph{shared} steady component using the best-sampled early epochs, then hold it fixed to track the flare component across all three epochs.
We model each SSA component with the standard homogeneous form
\begin{equation}
S(\nu)=S_p\left(\frac{\nu}{\nu_p}\right)^{5/2}
\left[1-\exp\!\left(-\tau(\nu)\right)\right],
\qquad
\tau(\nu)=\left(\frac{\nu}{\nu_p}\right)^{-(p+4)/2},
\label{eq:ssa_shape}
\end{equation}
and fit the total spectrum as $S_{\rm tot}=S_{\rm steady}+S_{\rm flare}$ using an MCMC sampler (\texttt{emcee}; \citealt{2013PASP..125..306F}) simultaneously across the three epoch-binned connected-array spectrum. The steady component shares $(S_p,\nu_p)$ across epochs, whereas the flare component has epoch-dependent $(S_p,\nu_p)$. We adopt $p=3$ for both components and include a 20\% fractional systematic uncertainty to capture non-simultaneity and calibration scatter. The best-fit parameters are reported in Table~\ref{tab:ssa_components}.

\subsection{Turnover-based equipartition scalings} \label{app:eq}
For a homogeneous SSA component with electron index $p \simeq 3$, the equipartition radius and magnetic field can be robustly estimated from the turnover frequency $\nu_p$ and peak flux density $S_p$ \citep[e.g.,][]{1994ApJ...426...51R, 1998ApJ...499..810C}. Using the convenient normalizations in \citet{1998ApJ...499..810C},
\begin{equation}
    R_{\rm eq} \simeq 4.0 \times 10^{14} \left(\frac{f}{0.5}\right)^{-1/19} \left(\frac{S_p}{\rm Jy}\right)^{9/19} \left(\frac{D_L}{\rm Mpc}\right)^{18/19} \left(\frac{\nu_p}{5\rm~GHz}\right)^{-1} (1+z)^{-1} {\rm ~cm},
\end{equation}
\begin{equation}
    B_{\rm eq} \simeq 1.1 \left(\frac{f}{0.5}\right)^{-4/19} \left(\frac{S_p}{\rm Jy}\right)^{-2/19} \left(\frac{D_L}{\rm Mpc}\right)^{-4/19} \left(\frac{\nu_p}{5\rm~GHz}\right) (1+z) {\rm ~G}.
\end{equation}
The resulting $R_{\rm eq}$ and $B_{\rm eq}$ values for the steady component and for each flare epoch are summarized in Table~\ref{tab:ssa_components}. Crucially, these estimates are remarkably resilient against microphysical uncertainties. Because the filling factor $f$ and the ratio of relativistic electron energy to magnetic energy enter the equations with very weak fractional powers ($-1/19$ and $-4/19$), even moderate deviations from strict equipartition do not alter the order-of-magnitude physical scales. Furthermore, while unknown Doppler boosting and specific geometric models affect the absolute normalizations, they only weakly impact the relative structural ratios between the steady and flare components, which securely drive the inferred density slope in Appendix~\ref{app:CNM}.

\subsection{Cooling and expansion timescales} \label{app:cooling}
The synchrotron cooling time for electrons emitting predominantly at observing frequency $\nu$ is
\begin{equation}
    t_{\rm syn} \approx 580 {\rm ~d~} \left(\frac{B}{1\rm~G}\right)^{-3/2} \left(\frac{\nu}{1\rm~GHz}\right)^{-1/2} (1+z)^{1/2}.
\end{equation}
At the peak of the high-frequency outburst (Epoch II), our equipartition estimate yields $B_{\rm eq} \approx 1.35$~G. This implies a characteristic synchrotron cooling time of $t_{\rm syn}(15{\rm~GHz}) \approx 1.0 \times 10^2$~days. In contrast, the adiabatic expansion timescale is $t_{\rm ad} \sim R_{\rm eq}/v \approx 3$~days, assuming an initial outflow velocity of $v = 0.2c$ at $R_{\rm eq} = 4.53 \times 10^{-4}$~pc.

The significant inequality $t_{\rm ad}\ll t_{\rm syn}$ is a useful sanity check: at GHz frequencies and $B\sim{\rm G}$, synchrotron cooling is slow, so adiabatic dilution and the time-dependence of particle acceleration dominate the light-curve shape.  If we took the equipartition size and assumed a single impulsive ejection expanding at $v\sim0.2\, c$, the adiabatic timescale would be only a few days, much shorter than the months-long high-frequency activity.  This does \emph{not} strictly exclude expansion, but it does imply that the emitting region cannot behave as a freely expanding one-shot plasmon.  The observations instead favor (i) substantially slower effective expansion due to confinement/strong deceleration in the dense CNM, (ii) sustained or repeated energy injection and in-situ particle acceleration (e.g., a long-lived outflow-driven shock), or (iii) a geometry in which $R_{\rm eq}$ underestimates the characteristic scale of the radiating volume.

\subsection{Apparent-speed limits from VLBI} \label{app:speed}

We constrain the apparent velocity ($\beta_{\rm app} \equiv v/c$) of any putative jet knot using two independent geometric limits over the $\Delta t_{\rm obs} \simeq 1.7$\,yr observational baseline (15.2 GHz observations from May 2022 to February 2024). At the redshift of J1430+2303 ($z=0.08105$), $1$\,mas corresponds to a projected physical scale of $\approx 1.526$\,pc, and the source-frame time interval is $\Delta t_{\rm src} = \Delta t_{\rm obs}/(1+z) \simeq 1.57$\,yr. The apparent proper motion speed is thus given by:
\begin{equation}
    \beta_{\rm app} = \frac{\Delta\theta\,(1.526\ {\rm pc\,mas^{-1}})}{\Delta t_{\rm src}\,(0.3066\ {\rm pc\,yr^{-1}}c^{-1})}.
    \label{eq:betaapp}
\end{equation}

First, the strict non-detection of a secondary component constrains the resolvable separation of a hypothetical moving knot. The practical threshold for distinguishing a secondary knot from the primary core can be approximated by:
\begin{equation}
    \Delta\theta_{\rm lim} \sim \max\!\left(0.3\text{--}0.5\,\theta_{\rm beam,min},\ 3\sigma_{\rm pos}\right), \quad \sigma_{\rm pos} \approx \frac{\theta_{\rm beam}}{2\,{\rm S/N}}.
\end{equation}
Given our 15.2\,GHz restoring beam minor axis of $\theta_{\rm beam,min} \sim 0.5$\,mas and typical signal-to-noise ratios of S/N $\sim 5$--$9$, the minimum resolvable separation limit falls in the range of $\Delta\theta_{\rm lim} \approx 0.15$--$0.25$\,mas. Substituting this angular limit into Equation~\ref{eq:betaapp} yields an apparent speed upper limit of $\beta_{\rm app} \lesssim 0.47$--$0.79$, conservatively capping the bulk proper motion at $\lesssim 0.5$--$0.8\,c$.

Independently, we can place a tighter illustrative constraint based purely on the structural size upper limits of the core, rather than searching for a resolved centroid shift. The rigorous MCMC-derived structural upper limits at 15.2\,GHz are $\theta < 0.22$\,mas in May 2022 and $\theta < 0.27$\,mas in February 2024. This restricts the maximum unresolvable expansion of the emitting region to $\Delta\theta \lesssim 0.05$\,mas over this period. Converting this differential size limit to a radial expansion speed yields an even stricter illustrative bound of $\beta_{\rm app} \lesssim 0.16$. Qualitatively, this stationary scenario is further corroborated by the multi-epoch relative astrometry (Fig.~\ref{fig:positions}), which reveals no obvious monotonic drift of the core centroid across the observing campaign. Both geometric arguments consistently confirm the highly compact and fundamentally sub-relativistic nature of the radio core.

\section{CNM density profile from the radio spectrum}\label{app:CNM}

The steady+flare SSA decomposition provides two characteristic radii: a larger-scale, persistent steady component and a smaller-scale, time-variable flare component. If the flare component traces an outflow/shock propagating through the same circumnuclear medium (CNM) responsible for the steady component, the pair $(R_{\rm eq},B_{\rm eq})$ can be mapped to an effective ambient density. For a strong shock with magnetic energy fraction $\epsilon_B$,
\begin{equation}
n \simeq \frac{B^2}{9\pi\,\epsilon_B\,m_p\,v^2}.
\end{equation}
Using $v=0.2c$ and $\epsilon_B=0.1$, the steady component ($R_{\rm eq}\simeq9.1\times10^{-3}$~pc; $B_{\rm eq}\simeq 0.1$~G) corresponds to $n\sim 60$~cm$^{-3}$, while the highly compact flare component implies a significantly denser environment of  $n\sim(5$--$11)\times10^3$~cm$^{-3}$ at $R\sim(4$--$6)\times10^{-4}$~pc (Table~\ref{tab:ssa_components}). Fitting a power-law density profile $n(R) \propto R^{-k}$ between these two distinct emission zones yields an index of $k \simeq 1.5$--$1.7$ across Epochs~I--III, revealing a steep inner CNM cusp. Crucially, this inferred slope is exceptionally robust. Because any systematic corrections to the absolute equipartition normalization (e.g., from specific geometric filling factors, microphysics, or modest Doppler boosting) shift both components by identical fractional amounts, the relative scaling is completely preserved. Thus, while the absolute density normalization carries order-of-magnitude uncertainties, the existence of the steep $1.5$--$1.7$ density gradient is a strict, model-independent consequence of the steady+flare spectral decomposition.

\end{appendix}

\end{document}